\documentclass[preprint]{aastex631}
\usepackage{color}
\newcommand{\B}[1]{{\color{blue}{#1}}}
\newcommand{\R}[1]{{\color{red}{#1}}}
\newcommand{\M}[1]{{\color{magenta}{#1}}}
\definecolor{DarkGreen}{rgb}{0.0,0.4,0.0}
\newcommand{\G}[1]{{\color{DarkGreen}{#1}}}
\usepackage[normalem]{ulem}
\usepackage{soul}
\usepackage{cancel}

\newcommand{\speed}[1]{#1 km\,s$^{-1}$}

\begin{document}
\setlength{\parskip}{10pt}
\title{High-Resolution Observation of Solar Prominence Plumes Induced by Enhanced Spicular Activity}



\author[0000-0002-9865-5245]{Wensi Wang}
\affiliation{CAS Key Laboratory of Geospace Environment, Department of Geophysics and Planetary Sciences\\
University of Science and Technology of China, Hefei, 230026, China}
\affiliation{Mengcheng National Geophysical Observatory\\
University of Science and Technology of China, Hefei, 230026, China}

\author[0000-0003-4618-4979]{Rui Liu}
\affiliation{CAS Key Laboratory of Geospace Environment, Department of Geophysics and Planetary Sciences\\
University of Science and Technology of China, Hefei, 230026, China}
\affiliation{Mengcheng National Geophysical Observatory\\
University of Science and Technology of China, Hefei, 230026, China}

\author[0000-0002-2559-1295]{Runbin Luo}
\affiliation{CAS Key Laboratory of Geospace Environment, Department of Geophysics and Planetary Sciences\\
University of Science and Technology of China, Hefei, 230026, China}

\author[0000-0003-2891-6267]{Xiaoli Yan}
\affiliation{Yunnan Observatories, Chinese Academy of Sciences, Kunming, 650011, China}
\affiliation{Yunnan Key Laboratory of Solar Physics and Space Science, Kunming, 650216, China}

\correspondingauthor{Rui Liu; Wensi Wang}
\email{rliu@ustc.edu.cn; minesnow@ustc.edu.cn}



\begin{abstract}
Solar prominences are the most prominent large-scale structures observed above the solar limb in emission in chromospheric lines but in absorption in coronal lines. At the bottom of prominences often appears a bubble, with plumes occasionally rising from the prominence-bubble interface. The plumes may potentially play an important role in the mass supply and thermodynamic evolution of prominences, but their nature and generation mechanism are elusive. Here we use the high-resolution H$\alpha$ observations obtained by the New Vacuum Solar Telescope (NVST) to investigate a quiescent prominence with bubbles and plumes on 8 November 2022. Within an interval of about two hours, enhanced spicular activity disturb the prominence-bubble interface, producing bursts of small-scale plumes rising through the prominence. Characterized by clustered spicules jetting at higher speeds (sometimes exceeding the typical chromopsheric Alfv\'{e}n speed) and longer life-time (over 15 minutes), the enhanced spicular activity differs markedly from regular spicules. We hence conjecture that the enhanced spicular activity may drive shock waves, which trigger the magnetic Richtmyer-Meshkov instability at the prominence-bubble interface, leading to the formation of small-scale plumes. These observations provide evidence that the enhanced spicular activity plays a potentially important role in the dynamic evolution of bubbles and plumes, thereby participating in the mass supply of solar prominences. 

\end{abstract}


\section{Introduction} \label{sec:intro}
Solar prominences are cool, dense “clouds” suspended in the hot, tenuous corona. In H$\alpha$ observations, prominences appear in emission above the limb, but in absorption against the disk (also termed filaments in this case, but often used interchangeably with prominences); in extreme ultraviolet (EUV) wavelengths, they generally have a dark appearance, but may brighten from time to time due to heating, e.g., at the interface region between the prominence and the ambient corona often emits in UV/EUV \citep{Labrosse2010physics}. The helical flux-rope model compares favorably with the prominence–cavity system, in which the prominence is surrounded by the cavity, an elliptical region of closed loops in limb observations \citep{Gibson2018solar}. Alternatively, a steady-state dynamic solution in a flat-topped sheared arcade can explain the formation and suspension of prominences through thermal non-equilibrium \cite[]{Karpen2005prominence}.

An arched, apparently transparent structure is sometimes observed beneath quiescent prominences, known as a prominence bubble. Plumes often form at the boundary between the bubble and the prominence, rising upward through the prominence. It is believed that prominence bubbles are caused by the emergence of magnetic flux from below the quiescent prominence \citep{Berger2017quiescent}; some may even possess a kinked flux rope configuration \citep{Awasthi2019mass}. The upflow plumes can be generated by the magnetic Rayleigh-Taylor  instability (RTI) due to the density inversion  \citep{Berger2010quiescent, Ryutova2010observation, Hillier2011numerical, Hillier2012simulations, Hillier2012numerical, Keppens2015solar} or by the coupled Kelvin–Helmholtz (KH)-RT instability due to localized shear flows at the prominence-bubble boundary \citep{Berger2017quiescent,Awasthi2019mass}. Alternatively, it is suggested that magnetic reconnection at the bubble boundary may drive the plume flows \citep{Dudik2012magnetic,Gunar2014magnetic,Shen2015fine}. Prominence bubbles and plumes are considered to play an important role in the mass supply and evolution of quiescent prominences \cite[]{Berger2011magneto,Berger2017quiescent}. 

To determine the physical mechanisms responsible for the formation and evolution of bubbles and plumes, high-resolution observations are indispensable, because of their small scales and highly dynamic nature. More recent studies employing the New Vacuum Solar Telescope \citep[NVST;][]{Liu2014nvst,Yan2020research} have provided some new insights into this topic. It has also been found that bubbles may collapse into large plumes under the impact of erupting minifilaments \cite[]{Chen2021solar,Wang2022formation,Guo2024formation}. The plume fronts are found to be associated with enhanced brightening, Doppler shifts, and nonthermal velocities, which seems to be consistent with the coupled KH-RT instability \cite[]{Xue2021observations}. Combining the limb and on-disk observations as well as NLFFF extrapolations, \cite{Guo2021reconstructing} proposed that the prominence-bubble boundary may corresponds to the interface between the prominence magnetic dips and the underlying magnetic loops.

In this paper, we use NVST observations to investigate a quiescent prominence with bubbles and plumes on 8 November 2022. We present how the impact of enhanced spicular activity on the boundary of a bubble drives numerous small-scale plumes rising through the prominence. Enhanced spicular activities are found to be associated with flux emergence or cancellation, suggesting that magnetic reconnection may be the underlying mechanism \cite[]{Samanta2019generation}. They are arguably mini-version of coronal jets in the chromosphere, probably driven by erupting microfilaments \cite[]{Sterling2020possible}. Unlike previous observations of prominence plumes, most of the plumes associated with enhanced spicular activity in the present case are smaller and have a blob shape, sometimes dragging a tail like tadpoles. In the sections that follow, the observation is described and analyzed in detail in Section~\ref{sec:obs} and Section~\ref{sec:res}. Concluding remarks are given in Section~\ref{sec:dis}.

\section{Overview of observations} \label{sec:obs}

\begin{figure}
    \centering
    \includegraphics[width=0.9\linewidth]{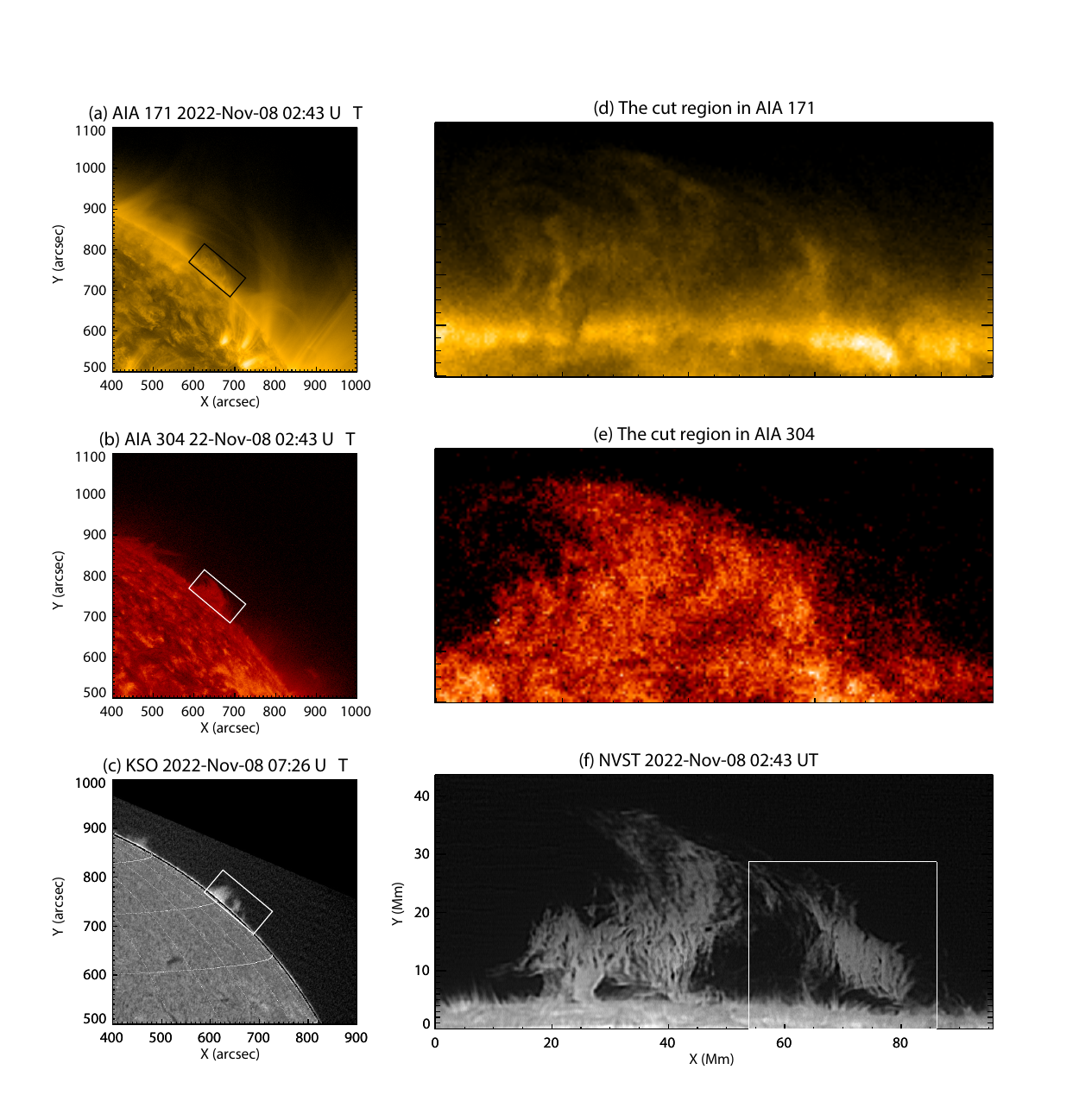}
    \caption{Overview of AIA and H$\alpha$ observations. (a) the AIA 171~{\AA} image. (b) the AIA 304~{\AA} (c) the H$\alpha$ image. The white box in these panels outlines the location of the prominence. (d) the cut region in AIA 171~{\AA}. (e) the cut region in AIA 304~{\AA}. (f) the NVST H$\alpha$ line center image. The white box in (f) shows the FOV of Figure~\ref{fig2}. An animation of NVST H$\alpha$ line center images is available. The animation proceeds from 02:56 UT to 06:59 UT on 2022 November 08, encompassing the complete evolution of the prominence. } 
    \label{fig1} 
\end{figure}

The quiescent prominence was observed above the northwestern limb on 08 November 2022 by the NVST, which is an 1-meter ground-based solar telescope located at the Fuxian Lake in Yunnan Province, China. The NVST provides high resolution H$\alpha$ (6562.8~{\AA}) filtergrams with a nominal temporal cadence of 12~s and a pixel size of $0''.165$. The temporal cadence in this observation is approximately 1 minute as we observed at four wavelength positions at the H$\alpha$ line wing ($\pm0.3$~{\AA} and $\pm0.5$~{\AA}) in addition to the H$\alpha$ line center. The observational period lasted for about 4 hours, from 02:43 UT to 07:00 UT, except for a data gap of about 15 minutes due to a temporary loss of the target. The prominence was also captured in EUV by the Atmospheric Imaging Assembly \citep[AIA;][]{Lemen2012aia} onboard the Solar Dynamics Observatory and by the H$\alpha$ telescope operated by the Kanzelhöhe Solar Observatory (KSO). The AIA and KSO images are co-aligned with the NVST observations according to the prominence features. Figure~\ref{fig1} shows the H$\alpha$ and EUV observations of the prominence. The NVST's field of view is around $126''\times126''$, which is large enough to cover the whole prominence (Figure~\ref{fig1}(f)). 

The prominence in the high-resolution H$\alpha$ observations takes a hedgerow appearance, consisting of myriads of vertical threads and exhibiting highly dynamic behaviors (see the animation accompanying Fig.~\ref{fig1}). It has a large dark void in the center, where diffuse prominence material is still present in the corresponding AIA 171 and 304~{\AA} images (Fig.~\ref{fig1}). To the left of the dark void, one can see cyclonic circular motions in the plane of sky, reminiscent of earthly tornadoes \cite[]{Li2012solar}; above the dark void, one can see bidirectional flows most likely along the spine of the prominence. Three bubbles are observed during the whole observational period. An expanding bright arc is briefly present in the dark void in the center of the prominence and soon collapse, associated with a lot of falling `fingers'. One bubble to the left of the dark void was low lying and its evolution is obscured by shrub-like spicules. Fortunately, the formation and evolution of the bubble to the right of the dark void is clearly captured by NVST. In this study, we will focus on the dynamic evolution of this bubble.

\section{Results}\label{sec:res}
\subsection{Dynamic evolution of the bubble of interest} 
\begin{figure}
    \centering
    \includegraphics[width=0.9\linewidth]{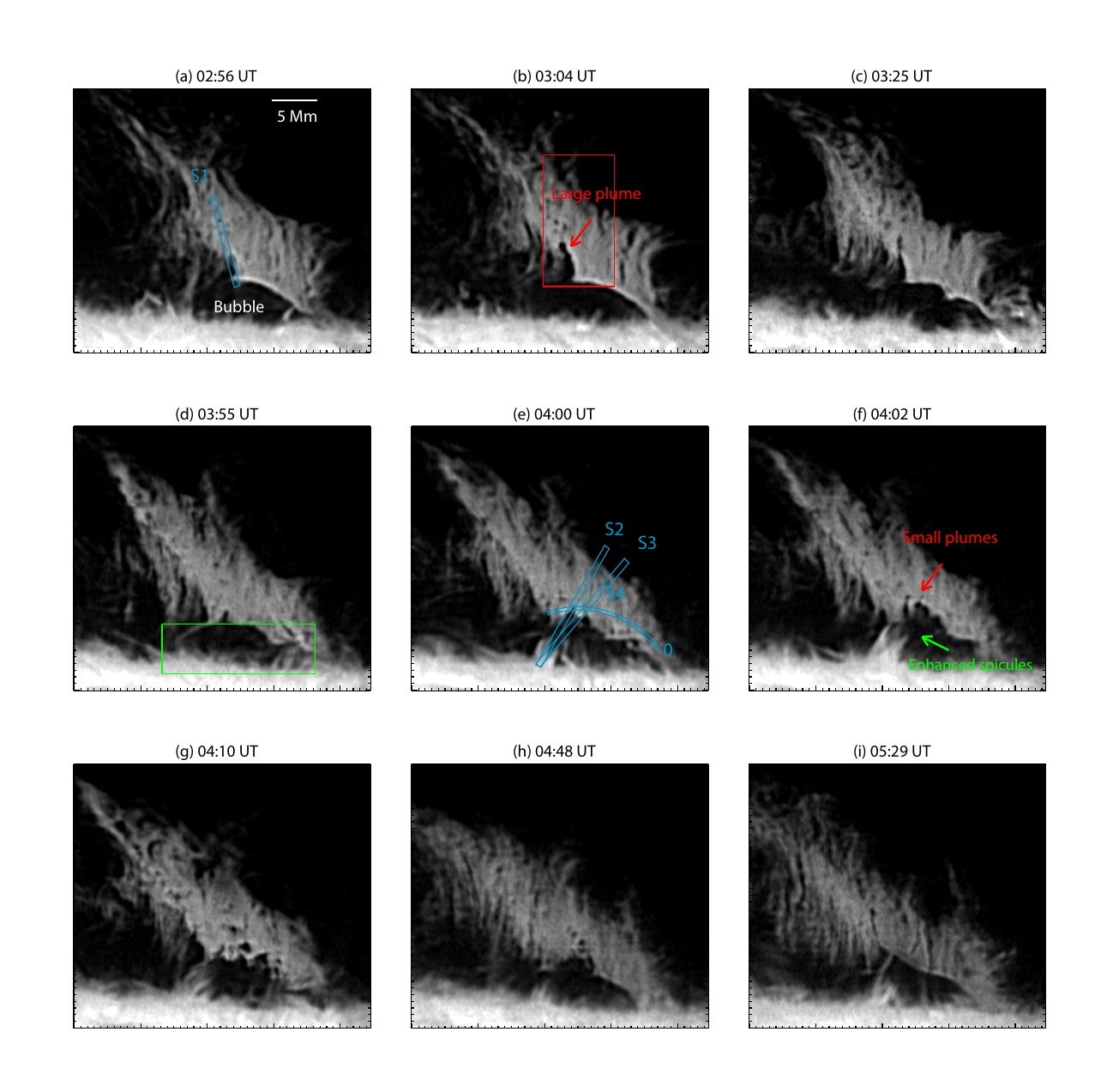}
    \caption{Impact of the enhanced spicules on the prominence-bubble interface. Snapshots show the temporal evolution of the bubble and plumes in the H$\alpha$ line center. The virtual slits are marked in blue in Panels (a) and (e). The red box in (b) shows the FOV of Figure~\ref{fig3}, and a red arrow marks a large plume. The green box in (d) shows the FOV of Figure~\ref{fig4}. Panels (e--g) show the 1st episode of impact. The red arrows mark the small plumes. The green arrows in the panels mark the enhanced spicules. Panels (h--i) show another two spicules as part of the enhanced spicular activity. An animation of H$\alpha$ line center images with the same field of view is available online. The animation proceeds from 02:56 UT to 06:59 UT on 2022 November 08, encompassing the complete evolution of the bubble.}
    \label{fig2} 
\end{figure}

Figure~\ref{fig2} and the accompanying animation show the complete evolution of the bubble to the right of the large dark void in the center of the prominence. In the beginning of the observation, the bubble region takes the appearance of a dark, semicircular void. Its right half part is separated from the vertical prominence threads by a bright arc spanning roughly a quarter of a circle (Figure~\ref{fig2}(a)), yet the left half part was not clearly defined. At about 03:04 UT, a sizable plume appears at the top of the bubble (Figure~\ref{fig2}(b)). 

Meanwhile, the bright arc is under perturbation (Figure~\ref{fig2}(c)) and ultimately disintegrates around 03:40 UT (see the animation accompanying Fig.~\ref{fig2}). Within the subsequent two hours, enhanced spicular activity impact the interface between the prominence and the bubble, resulting in a `storm' of small-scale plumes at the interface (\S\ref{subsec:impact}). Then after the temporary loss of the target from 05:43 UT to 06:00 UT, the bubble is observed to expand, and possesses a bright semicircle boundary from about 06:10 UT onward.

\subsection{Formation of small-scale plumes} \label{subsec:impact}
The most prominent episode of enhanced spicular activity (hereafter `ES1'), which consists of a group of spicules and reaches higher altitudes than regular ones, is detailed in Figure~\ref{fig2}(e--g). At about 04:00 UT, a group of clustered spicules appear inside the bubble below the prominence and stream toward the interface. About 2 minutes later, as soon as the spicules hit the interface (Figure~\ref{fig2}(f)), a burst of small plumes with tadpole shapes and dark appearances form immediately above the interface, rising into the prominence. In the subsequent five minutes, as the clustered spicules continue to impact the interface (see the animation accompanying Fig.~\ref{fig2}), smaller plumes continue to form and ascend from the interface. One particular plume thins progressively as it rises through the prominence, while other plumes often split into even smaller plumes (Figure~\ref{fig2}(g)). The majority of the clustered spicules have dissipated by 04:08 UT, except one spicule which is still apparently connected to the prominence (see Figure~\ref{fig2}(g) and the accompanying animation), suggesting that the group of clustered spicules do not impact the prominence simultaneously but persist for a short time period in an intermittent manner. A subsequent episode of clustered spicules (hereafter `ES2') emerges at about 05:30 UT in the similar location, but generally proceeds in a less intense manner than ES1. Simultaneously, another large plume seems to be in formation right above ES2, associated with a few small-scale plumes apparently torn from its top (see the animation accompanying Figure~\ref{fig2}).

\begin{figure}
    \centering
    \includegraphics[width=0.9\textwidth]{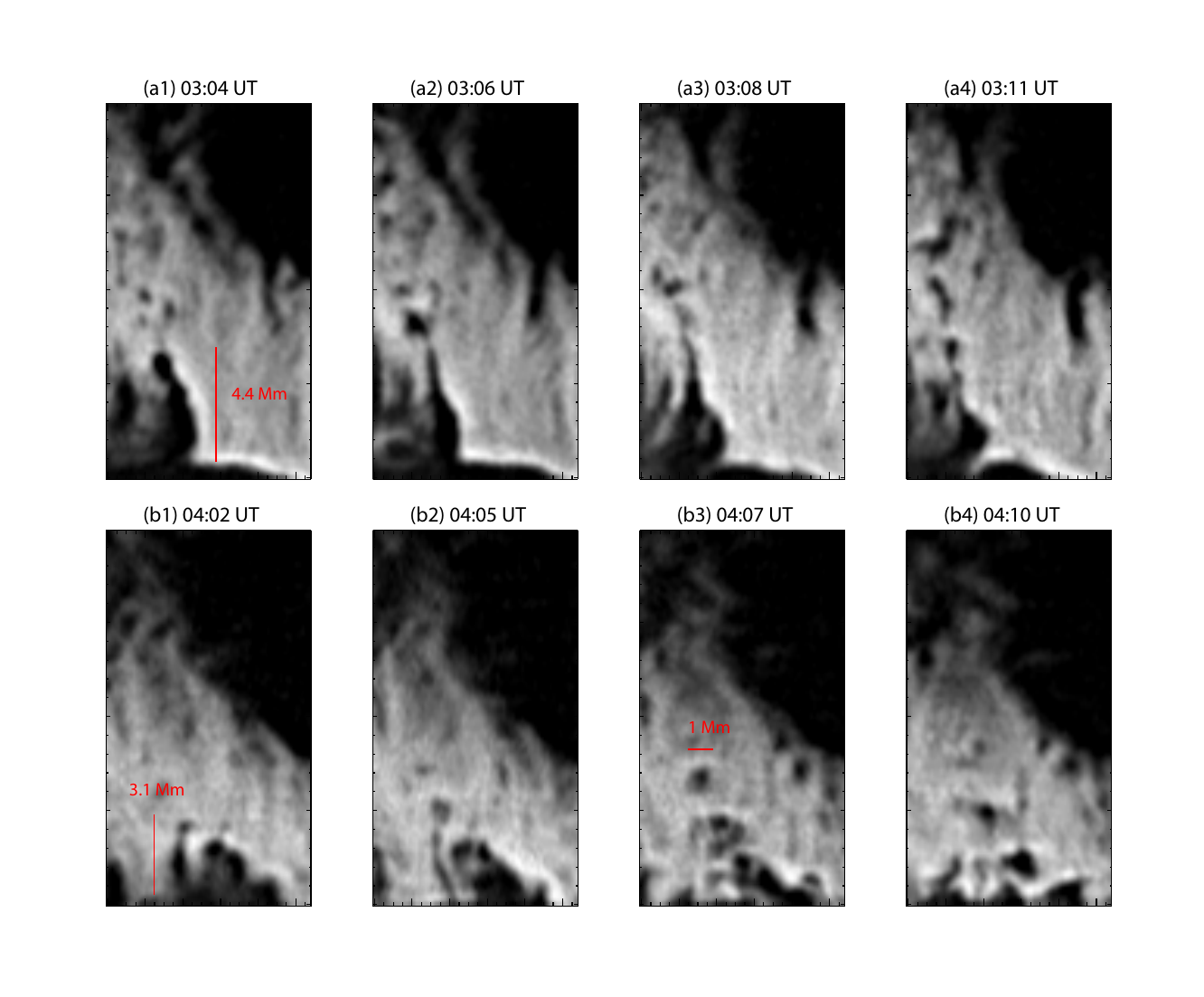}
    \caption{Evolution of two different types of plumes. Panels (a1--a4) show that a large-scale plume breaks into several smaller plumes. Panels (b1--b4) show a burst of small-scale plumes forming at the prominence-bubble interface. An animation of H$\alpha$ line center images is available, showing the evolution of the large-scale plume (left) side by side with that of small-scale plumes (right). The animation proceeds from 03:00 UT to 03:40 UT on 2022 November 08 on the left and from 04:00 to 04:40 UT on 2022 November 08 on the right.}
    \label{fig3} 
\end{figure}

\begin{figure}
    \centering
    \includegraphics[width=0.9\linewidth]{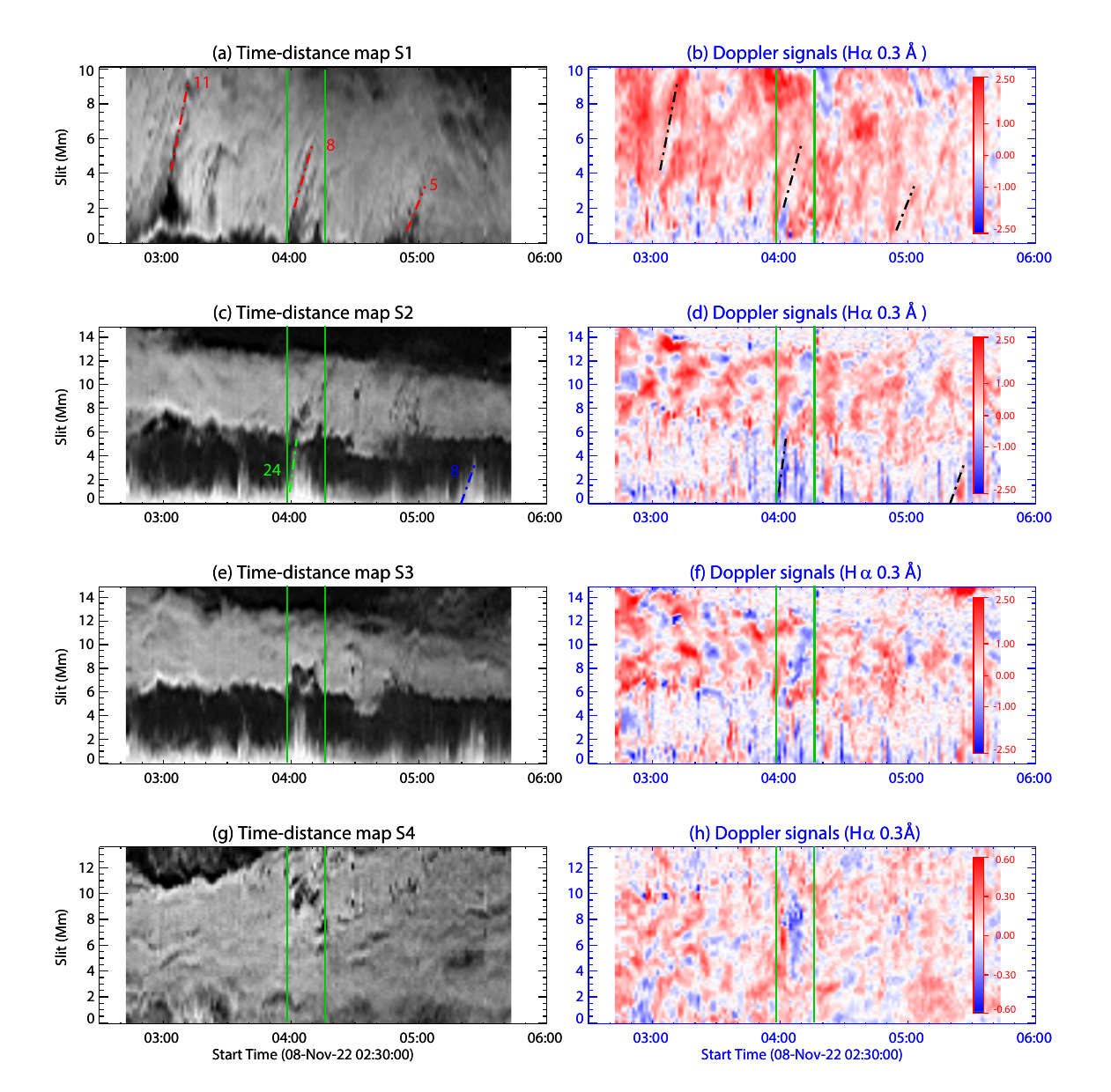}
    \caption{Time-distance maps and corresponding Doppler maps constructed from various virtual slits. The slits are marked in Figure~\ref{fig2}. In the left column, the dot-dashed lines indicate the trajectories of plumes (red) and of enhanced spicules ES1 (green) and ES2 (blue), and in the right column they are reproduced in black. The green vertical lines mark the interval of ES1, same as in Figure~\ref{fig5}(a). }
    \label{fig4} 
\end{figure}

\begin{figure}
    \centering
    \includegraphics[width=0.9\linewidth]{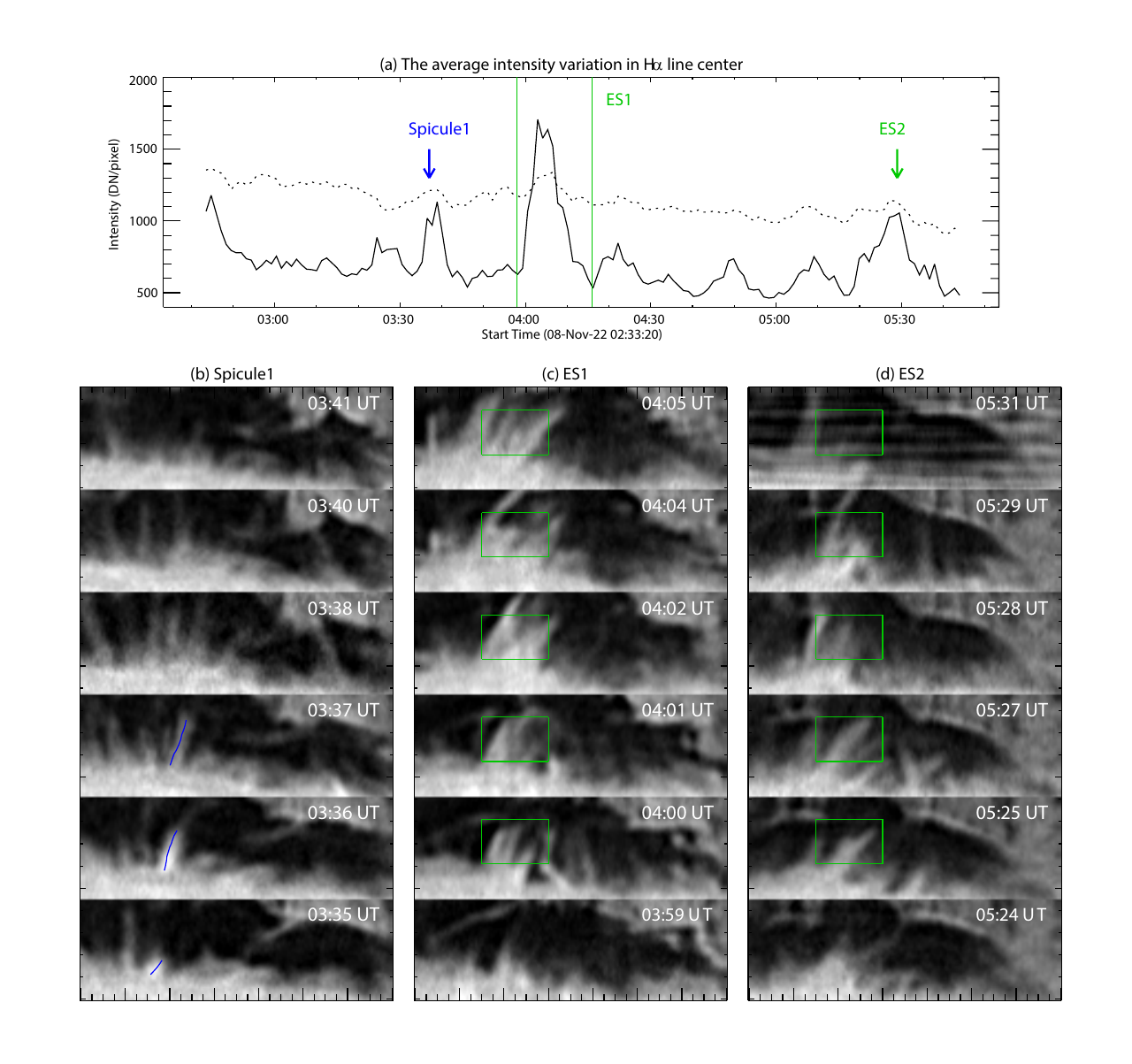}
        \caption{Comparison between regular and enhanced spicules. All snapshots are from H$\alpha$ line center images. Panel (a) shows the evolution of the average intensities in the H$\alpha$ line center. The solid curve represents the average intensity computed within the green box in (c) and (d), while the dotted one indicates the mean intensity within the field of view of the snapshots. The blue line in (b) delineates one regular spicule (Spicule1), and the green box in (c) and (d) brings attention to two episodes of enhanced spicular activity (ES1 and ES2). } 
    \label{fig5}
\end{figure}

The characteristics of these small-scale plumes, typically with a vertical extent less than 3 Mm and a horizontal extent less than 1 Mm, are distinct from those of a large plume (Fig.~\ref{fig2}b), which appears to be spontaneously formed prior to the ejection of clustered spicules and has a size typically a few times larger than the small-scale plumes. Figure~\ref{fig3} and accompanying animation compare the plumes formed before and after the appearance of ES1. To further clarify the speed and direction of the plume movement, we apply four virtual slits with distinct orientations to H$\alpha$ images. The slit S1 is oriented along the movement trajectory of the large plume; S2 follows the direction of enhanced spicular activity; S3 deviates from S2 by 10 deg; S4 is curved along the prominence-bubble interface, cutting across both the large plume and small-scale plumes. The resultant time-distance stack plots are shown in Figure~\ref{fig4}. The rising of the large plume at about 03:00 UT is tracked by S1, giving a speed about \speed{11} (Fig.~\ref{fig4}(a)). The base of the large plume was kept attached to the prominence-bubble interface for about 20 minutes, during which its top was intermittently torn into smaller plumes rising upward through the prominence before eventually mixing into the prominence (Figure~\ref{fig3}(a1)-(a3)). Then the large plume rapidly disintegrates (Figure~\ref{fig3}(a4)). About 50 minutes later, the small plumes are observed to initiate at the prominence-bubble interface and rise through the prominence, following the impact of enhanced spicular activity. Rising with speeds ranging from 4 to \speed{10}, these small plumes are released from the interface over a wide angular range, as seen through the virtual slits (S1--S3) which are oriented in different directions, during the period of ES1 (marked by green vertical lines in Figure~\ref{fig4} \& \ref{fig5}). Furthermore, no obvious plumes were observed prior to the enhanced spicular activity in the S2 and S3 directions. In contrast to the large plume, which has a lifetime exceeding 20 min, the small-scale plumes fragment into even smaller plumes within 5 minutes. 

This type of enhanced spicular activity is markedly distinct from the regular spicules, as compared side by side in Figure~\ref{fig5}. Figure~\ref{fig5}b shows an exemplary spicule (Spicule1), which is relatively isolated. In contrast, the enhanced spicular activity manifests as a group of spicules suddenly appearing collectively and lasts for more than 15 minutes. In particular, ES1 at about 04:00 UT (Figure~\ref{fig5}c) has significantly higher peak intensity than Spicule1 (Figure~\ref{fig5}a). ES2 (Figure~\ref{fig5}d) at about 05:25 UT is less intense than ES1, displaying a similar peak intensity as Spicule1, but ES2's period of enhanced intensity lasts as long as ES1. ES2 has a slower speed ($\sim\,$8 km~s$^{-1}$) and reaches lower heights than ES1 (Figure~\ref{fig4}c), which is probably why it fails to induce bursts of small-scale plumes as ES1. The time-distance stack plots derived from the slit S2 aligned along the enhanced spicules (Figure~\ref{fig2}(c)) further demonstrate the close spatio-temporal association between the enhanced spicular activity and small-scale plumes. One can see that the small-scale plumes appear immediately as the enhanced spicules reach the height of the prominence-bubble interface (Figure~\ref{fig4}e). A caveat to keep in mind, however, is the possibility that the spicules might apparently reach the prominence height in projection but are actually located in the foreground/background of the bubble.

To investigate the line-of-fight motions inside the prominence and the enhanced spicules, we estimated the Doppler shifts by the following formula:
\[D=\frac{I_{b}-I_{r}}{I_{b}+I_{r}}\]
where $I_{b}$ and $I_{r}$ refer to the intensities in the H$\alpha$ blue and red wings, respectively. The sign of $D$ represents the Doppler signal, but the magnitude of the numerical value indicates only the strength of the observed signal, not the magnitude of the Doppler velocity. We calculate the Doppler shifts using two pairs of H$\alpha$ line wing observations ($\pm0.3$~{\AA} and $\pm0.5$~{\AA}) and obtained qualitatively similar results. We show Doppler signals from $\pm0.3~{\AA}$ in Figure~\ref{fig4}. The prominence and the large plume exhibit mixed Doppler signals. Seen through the virtual slits (Fig.~\ref{fig4}(d, f, \& h)), the enhanced spicules and associated small plumes are both dominated by blue shifts. But plumes tracked by S1, which is not oriented along the direction of the enhanced spicules, exhibit dominantly red shifts (Figure~\ref{fig4}b). This suggests that the plumes that are not directly produced under the impact of the enhanced spicules may rise in different directions. 

\section{Discussion \& Conclusion}\label{sec:dis}
In this paper, we investigate the formation and evolution of plumes in the quiescent prominence on 2022 November 08. The high-resolution NVST observations reveal that as soon as enhanced spicular activities impact the interface between the prominence and the bubble, a storm of small plumes appear at the interface and rise upward through the prominence. The unprecedented observation presents a unique opportunity for us to understand the dynamic evolution of solar prominence. 

Previous studies proposed that the plumes are generated by the magnetic RTI, Or coupled KH-RT instability. Indeed, the large plume observed here forms without any obvious external drivers and evolves similarly as in previous observations \citep[e.g.,][]{Berger2010quiescent}. However, 3D MHD simulations have demonstrated that rising plumes that form through only the magnetic RTI are much less dynamic than falling fingers, with the dynamics of both features being regulated by the evolving vorticity source terms \citep{Jenkins2022resolving}. Under the impact of enhanced spicular activities, the abundance of rising plumes in contrast to the lack of falling fingers is consistent with the presence an initial, external driver, as already suggested by \cite{Berger2011magneto}. Compared with the large plumes that seem to form spontaneously at the bubble boundary \citep[e.g.,][]{Berger2010quiescent,Xue2021observations,Chen2021solar}, the small plumes produced by enhanced spicular activities, however, display distinct evolution and morphology, with slower rising speeds, shorter lifetimes, and blob-like shapes, similar to those split from the large plume. Further, unlike the erupting minifilaments that produce plumes large enough to disrupt the whole bubble \citep[]{Chen2021solar,Wang2022formation,Guo2024formation}, the enhanced spicular activity seems to disturb bubbles less disruptively, which suggests that the generation mechanism of the small plumes might be different. Particularly, when the clustered spicules shoot upward at a projected speed of 20--\speed{30} (Figure~\ref{fig4}), exceeding the typical Alfv\'{e}n speed in the quiescent chromosphere, they are potentially capable of driving shock waves, explaining that the storm of plumes are released from the prominence-bubble interface over a wide angular range. 

These observational aspects are reminiscent of the Richtmyer-Meshkow instability \cite[RMI;][]{Brouillette2002rm,Zhou2017}, which arises when a shock wave passes through an interface separating two different fluids. The RMI can be considered as the impulsive version of the RTI. A major distinction lies in the acceleration direction:  the RMI occurs no matter which side of the interface that the acceleration is directed toward, but the RTI only works for acceleration toward the lighter fluid. Hence, that the enhanced spicular activity impacts on the heavier prominence plasma above the interface from below, together with their close spatio-temporal correlation with small plumes rising though the prominence, argues strongly in favor of the RMI over the RTI, although the latter may also participate in the process. Additionally, the RMI is suggested to play a role in the occurrence of supper-arcade downflows (SADs) observed in the decay phase of solar flares \cite[]{Shen2022origin}. Indeed, the small plumes, sometimes dragging a tail to give a tadpole-like appearance, share morphological similarities with SADs. Since neither the storm of plumes nor the enhanced spicular activity is observed to occur frequently at prominence bubbles, their spatio-temporal association is unlikely a coincidence. However, the potential projection effects and the lack of one-on-one correspondence in the current observation make it difficult to exclude the possibility of coincidental occurrence of the two phenomena. Further investigations by carrying out a statistical study and by combining numerical modeling with high-resolution observations will help lift the ambiguity.

\begin{acknowledgments}
This work was supported by the NSFC (12373064, 42274204, 42188101, 11925302, and 12325303), the Strategic Priority Program of the Chinese Academy of Sciences (XDB0560102), and the National Key R\&D Program of China (2022YFF0503002) and the USTC Research Funds of the Double First-Class Initiative. 
\facilities{NVST, SDO, KSO}
\end{acknowledgments}

\bibliography{sample631}{}

\begin{thebibliography}{}
\expandafter\ifx\csname natexlab\endcsname\relax\def\natexlab#1{#1}\fi
\providecommand{\url}[1]{\href{#1}{#1}}
\providecommand{\dodoi}[1]{doi:~\href{http://doi.org/#1}{\nolinkurl{#1}}}
\providecommand{\doeprint}[1]{\href{http://ascl.net/#1}{\nolinkurl{http://ascl.net/#1}}}
\providecommand{\doarXiv}[1]{\href{https://arxiv.org/abs/#1}{\nolinkurl{https://arxiv.org/abs/#1}}}

\bibitem[{{Awasthi} \& {Liu}(2019)}]{Awasthi2019mass}
{Awasthi}, A.~K., \& {Liu}, R. 2019, Frontiers in Physics, 7, 218,
  \dodoi{10.3389/fphy.2019.00218}

\bibitem[{{Berger} {et~al.}(2017){Berger}, {Hillier}, \&
  {Liu}}]{Berger2017quiescent}
{Berger}, T., {Hillier}, A., \& {Liu}, W. 2017, \apj, 850, 60,
  \dodoi{10.3847/1538-4357/aa95b6}

\bibitem[{{Berger} {et~al.}(2011){Berger}, {Testa}, {Hillier}, {Boerner},
  {Low}, {Shibata}, {Schrijver}, {Tarbell}, \& {Title}}]{Berger2011magneto}
{Berger}, T., {Testa}, P., {Hillier}, A., {et~al.} 2011, \nat, 472, 197,
  \dodoi{10.1038/nature09925}

\bibitem[{{Berger} {et~al.}(2010){Berger}, {Slater}, {Hurlburt}, {Shine},
  {Tarbell}, {Title}, {Lites}, {Okamoto}, {Ichimoto}, {Katsukawa}, {Magara},
  {Suematsu}, \& {Shimizu}}]{Berger2010quiescent}
{Berger}, T.~E., {Slater}, G., {Hurlburt}, N., {et~al.} 2010, \apj, 716, 1288,
  \dodoi{10.1088/0004-637X/716/2/1288}

\bibitem[{{Brouillette}(2002)}]{Brouillette2002rm}
{Brouillette}, M. 2002, Annual Review of Fluid Mechanics, 34, 445,
  \dodoi{10.1146/annurev.fluid.34.090101.162238}

\bibitem[{{Chen} {et~al.}(2021){Chen}, {Su}, {Xue}, {Gan}, \&
  {Huang}}]{Chen2021solar}
{Chen}, C., {Su}, Y., {Xue}, J., {Gan}, W., \& {Huang}, Y. 2021, \apjl, 923,
  L10, \dodoi{10.3847/2041-8213/ac3bd0}

\bibitem[{{Dud{\'\i}k} {et~al.}(2012){Dud{\'\i}k}, {Aulanier}, {Schmieder},
  {Zapi{\'o}r}, \& {Heinzel}}]{Dudik2012magnetic}
{Dud{\'\i}k}, J., {Aulanier}, G., {Schmieder}, B., {Zapi{\'o}r}, M., \&
  {Heinzel}, P. 2012, \apj, 761, 9, \dodoi{10.1088/0004-637X/761/1/9}

\bibitem[{{Gibson}(2018)}]{Gibson2018solar}
{Gibson}, S.~E. 2018, Living Reviews in Solar Physics, 15, 7,
  \dodoi{10.1007/s41116-018-0016-2}

\bibitem[{{Gun{\'a}r} {et~al.}(2014){Gun{\'a}r}, {Schwartz}, {Dud{\'\i}k},
  {Schmieder}, {Heinzel}, \& {Jur{\v{c}}{\'a}k}}]{Gunar2014magnetic}
{Gun{\'a}r}, S., {Schwartz}, P., {Dud{\'\i}k}, J., {et~al.} 2014, \aap, 567,
  A123, \dodoi{10.1051/0004-6361/201322777}

\bibitem[{{Guo} {et~al.}(2021){Guo}, {Hou}, {Li}, \&
  {Zhang}}]{Guo2021reconstructing}
{Guo}, Y., {Hou}, Y., {Li}, T., \& {Zhang}, J. 2021, \apjl, 911, L9,
  \dodoi{10.3847/2041-8213/abee92}

\bibitem[{{Guo} {et~al.}(2024){Guo}, {Hou}, {Li}, {Shen}, {Wang}, {Zhang},
  {Zheng}, {Wang}, \& {Mei}}]{Guo2024formation}
{Guo}, Y., {Hou}, Y., {Li}, T., {et~al.} 2024, \apj, 970, 110,
  \dodoi{10.3847/1538-4357/ad54b8}

\bibitem[{{Hillier} {et~al.}(2012{\natexlab{a}}){Hillier}, {Berger}, {Isobe},
  \& {Shibata}}]{Hillier2012simulations}
{Hillier}, A., {Berger}, T., {Isobe}, H., \& {Shibata}, K. 2012{\natexlab{a}},
  \apj, 746, 120, \dodoi{10.1088/0004-637X/746/2/120}

\bibitem[{{Hillier} {et~al.}(2011){Hillier}, {Isobe}, {Shibata}, \&
  {Berger}}]{Hillier2011numerical}
{Hillier}, A., {Isobe}, H., {Shibata}, K., \& {Berger}, T. 2011, \apjl, 736,
  L1, \dodoi{10.1088/2041-8205/736/1/L1}

\bibitem[{{Hillier} {et~al.}(2012{\natexlab{b}}){Hillier}, {Isobe}, {Shibata},
  \& {Berger}}]{Hillier2012numerical}
---. 2012{\natexlab{b}}, \apj, 756, 110, \dodoi{10.1088/0004-637X/756/2/110}

\bibitem[{{Jenkins} \& {Keppens}(2022)}]{Jenkins2022resolving}
{Jenkins}, J.~M., \& {Keppens}, R. 2022, Nature Astronomy, 6, 942,
  \dodoi{10.1038/s41550-022-01705-z}

\bibitem[{{Karpen} {et~al.}(2005){Karpen}, {Tanner}, {Antiochos}, \&
  {DeVore}}]{Karpen2005prominence}
{Karpen}, J.~T., {Tanner}, S.~E.~M., {Antiochos}, S.~K., \& {DeVore}, C.~R.
  2005, \apj, 635, 1319, \dodoi{10.1086/497531}

\bibitem[{{Keppens} {et~al.}(2015){Keppens}, {Xia}, \&
  {Porth}}]{Keppens2015solar}
{Keppens}, R., {Xia}, C., \& {Porth}, O. 2015, \apjl, 806, L13,
  \dodoi{10.1088/2041-8205/806/1/L13}

\bibitem[{{Labrosse} {et~al.}(2010){Labrosse}, {Heinzel}, {Vial}, {Kucera},
  {Parenti}, {Gun{\'a}r}, {Schmieder}, \& {Kilper}}]{Labrosse2010physics}
{Labrosse}, N., {Heinzel}, P., {Vial}, J.~C., {et~al.} 2010, \ssr, 151, 243,
  \dodoi{10.1007/s11214-010-9630-6}

\bibitem[{{Lemen} {et~al.}(2012){Lemen}, {Title}, {Akin}, {Boerner}, {Chou},
  {Drake}, {Duncan}, {Edwards}, {Friedlaender}, {Heyman}, {Hurlburt}, {Katz},
  {Kushner}, {Levay}, {Lindgren}, {Mathur}, {McFeaters}, {Mitchell}, {Rehse},
  {Schrijver}, {Springer}, {Stern}, {Tarbell}, {Wuelser}, {Wolfson}, {Yanari},
  {Bookbinder}, {Cheimets}, {Caldwell}, {Deluca}, {Gates}, {Golub}, {Park},
  {Podgorski}, {Bush}, {Scherrer}, {Gummin}, {Smith}, {Auker}, {Jerram},
  {Pool}, {Soufli}, {Windt}, {Beardsley}, {Clapp}, {Lang}, \&
  {Waltham}}]{Lemen2012aia}
{Lemen}, J.~R., {Title}, A.~M., {Akin}, D.~J., {et~al.} 2012, \solphys, 275,
  17, \dodoi{10.1007/s11207-011-9776-8}

\bibitem[{{Li} {et~al.}(2012){Li}, {Morgan}, {Leonard}, \&
  {Jeska}}]{Li2012solar}
{Li}, X., {Morgan}, H., {Leonard}, D., \& {Jeska}, L. 2012, \apjl, 752, L22,
  \dodoi{10.1088/2041-8205/752/2/L22}

\bibitem[{{Liu} {et~al.}(2014){Liu}, {Xu}, {Gu}, {Wang}, {You}, {Shen}, {Lu},
  {Jin}, {Chen}, {Lou}, {Li}, {Liu}, {Xu}, {Rao}, {Hu}, {Li}, {Fu}, {Wang},
  {Bao}, {Wu}, \& {Zhang}}]{Liu2014nvst}
{Liu}, Z., {Xu}, J., {Gu}, B.-Z., {et~al.} 2014, Research in Astronomy and
  Astrophysics, 14, 705, \dodoi{10.1088/1674-4527/14/6/009}

\bibitem[{{Ryutova} {et~al.}(2010){Ryutova}, {Berger}, {Frank}, {Tarbell}, \&
  {Title}}]{Ryutova2010observation}
{Ryutova}, M., {Berger}, T., {Frank}, Z., {Tarbell}, T., \& {Title}, A. 2010,
  \solphys, 267, 75, \dodoi{10.1007/s11207-010-9638-9}

\bibitem[{{Samanta} {et~al.}(2019){Samanta}, {Tian}, {Yurchyshyn}, {Peter},
  {Cao}, {Sterling}, {Erd{\'e}lyi}, {Ahn}, {Feng}, {Utz}, {Banerjee}, \&
  {Chen}}]{Samanta2019generation}
{Samanta}, T., {Tian}, H., {Yurchyshyn}, V., {et~al.} 2019, Science, 366, 890,
  \dodoi{10.1126/science.aaw2796}

\bibitem[{{Shen} {et~al.}(2022){Shen}, {Chen}, {Reeves}, {Yu}, {Polito}, \&
  {Xie}}]{Shen2022origin}
{Shen}, C., {Chen}, B., {Reeves}, K.~K., {et~al.} 2022, Nature Astronomy, 6,
  317, \dodoi{10.1038/s41550-021-01570-2}

\bibitem[{{Shen} {et~al.}(2015){Shen}, {Liu}, {Liu}, {Chen}, {Su}, {Xu}, \&
  {Liu}}]{Shen2015fine}
{Shen}, Y., {Liu}, Y., {Liu}, Y.~D., {et~al.} 2015, \apjl, 814, L17,
  \dodoi{10.1088/2041-8205/814/1/L17}

\bibitem[{{Sterling} {et~al.}(2020){Sterling}, {Moore}, {Samanta}, \&
  {Yurchyshyn}}]{Sterling2020possible}
{Sterling}, A.~C., {Moore}, R.~L., {Samanta}, T., \& {Yurchyshyn}, V. 2020,
  \apjl, 893, L45, \dodoi{10.3847/2041-8213/ab86a5}

\bibitem[{{Wang} {et~al.}(2022){Wang}, {Yan}, {Xue}, {Yang}, {Li}, {Chen},
  {Xia}, \& {Liu}}]{Wang2022formation}
{Wang}, J., {Yan}, X., {Xue}, Z., {et~al.} 2022, \aap, 659, A76,
  \dodoi{10.1051/0004-6361/202142584}

\bibitem[{{Xue} {et~al.}(2021){Xue}, {Vial}, {Su}, {Li}, {Xu}, {Su}, {Zhou}, \&
  {Li}}]{Xue2021observations}
{Xue}, J.-C., {Vial}, J.-C., {Su}, Y., {et~al.} 2021, Research in Astronomy and
  Astrophysics, 21, 222, \dodoi{10.1088/1674-4527/21/9/222}

\bibitem[{{Yan} {et~al.}(2020){Yan}, {Liu}, {Zhang}, \& {Xu}}]{Yan2020research}
{Yan}, X., {Liu}, Z., {Zhang}, J., \& {Xu}, Z. 2020, Science in China E:
  Technological Sciences, 63, 1656, \dodoi{10.1007/s11431-019-1463-6}

\bibitem[{{Zhou}(2017)}]{Zhou2017}
{Zhou}, Y. 2017, \physrep, 720, 1, \dodoi{10.1016/j.physrep.2017.07.005}

\end{thebibliography}
\bibliographystyle{aasjournal}



\end{document}